\documentstyle[preprint,aps]{revtex}

\tightenlines
\newcommand{\be}{\begin{equation}}
\newcommand{\ee}{\end{equation}}
\newcommand{\bd}{\begin{displaymath}}
\newcommand{\ed}{\end{displaymath}}
\newcommand{\bea}{\begin{eqnarray}}
\newcommand{\eea}{\end{eqnarray}}

\newcommand{\bRt}{\mbox{${\bf R}(t)$}}
\newcommand{\bR}{\mbox{${\bf R}$}}
\newcommand{\bBt}{\mbox{${\bf B}(t)$}}
\newcommand{\bB}{\mbox{${\bf B}$}}
\newcommand{\PM}{\mbox{$P_{-}$}}
\newcommand{\PP}{\mbox{$P_{+}$}}
\newcommand{\TP}{\mbox{$T_{+}$}}
\newcommand{\TM}{\mbox{$T_{-}$}}

\newcommand{\cPP}{\mbox{${\cal P}_{+}$}}
\newcommand{\cPM}{\mbox{${\cal P}_{-}$}}

\newcommand{\terma}{\mbox{$\left[ g\tan\frac{\Omega_{0}t}{2} \right]$}}
\newcommand{\termb}{\mbox{$\tan\frac{\delta t}{2}$}}

\begin{document}
\draft

\title{Berry's Phase in the Presence of a Non-Adiabatic Environment}

\author{Frank Gaitan}
\address{Department of Physics; Boston College; Chestnut Hill, MA 02167-3811}
\date{\today}

\maketitle

\begin{abstract}
We consider a two-level system coupled to an environment that evolves
non-adiabatically. We present a non-perturbative method for determining
the persistence
amplitude whose phase contains all the corrections to Berry's phase produced
by the non-adiabatic motion of the environment. Specifically, it includes
the effect of transitions between the two energy levels to all orders
in the non-adiabatic coupling. The problem of determining all non-adiabatic
corrections is reduced to solving an ordinary differential equation to
which numerical methods should provide solutions in a 
variety of situations. We apply our method to a particular
example that can be realized as a magnetic resonance experiment, thus raising
the possibility of testing our results in the lab. 
\end{abstract}


\section{Introduction}
\label{sec1}
In the original Berry phase scenario \cite{br1}, the focus of attention is a
quantum system with a discrete, non-degenerate energy spectrum. Its
Hamiltonian $H[\bR]$ is assumed to depend on a set of classical parameters
$\bR$ which represent an environmental degree of freedom to which the
quantum system is coupled. The environment is assumed to evolve 
adiabatically. This produces an adiabatic time dependence in the quantum
Hamiltonian, $H=H[\bRt ]$. The time dependence of the quantum state
$|\psi (t)\rangle$ is determined by solving Schrodinger's equation using
the quantum adiabatic theorem. Towards this end, one introduces the energy
eigenstates of the instantaneous Hamiltonian $H[\bRt ]$,
\begin{displaymath}
H[\bRt ] |E[\bRt ]\rangle = E[\bRt ] |E[\bRt ]\rangle \hspace{0.1in} .
\end{displaymath}
It is further assumed that the environment is taken adiabatically around a
loop in parameter space such that $\bR (T) = \bR (0)$, and that the 
quantum system is initially prepared in an eigenstate $|E[\bR (0)] \rangle$
of the initial Hamiltonian $H[\bR (0)]$. The quantum adiabatic theorem
states that, at time $t$, the quantum system will be found in the state
$|E[\bRt ]\rangle$ to within a phase factor,
\be
|\psi (t) \rangle = \exp \left[\, i\gamma_{E}(t) -\frac{i}{\hbar}
                     \int_{0}^{t}\, d\tau\, E[\bR (\tau)] \right]
                      |E[\bRt ]\rangle \hspace{0.1in} .
\ee
The second term in the phase of the exponential  is known as the dynamical
phase and was already familiar from previous studies of the quantum 
adiabatic theorem. The first term represents Berry's discovery, and is 
referred to as Berry's phase,
\be
\gamma_{E}(t) = i\int_{0}^{t}\, d\tau\, \langle E[\bR (\tau )]|
                 \frac{\partial}{\partial\tau} |E[\bR (\tau ) ]\rangle
                  \hspace{0.1in} .
\ee
In the cases where Berry's phase is physically relevant, $\gamma_{E}$
is non-integrable: it cannot be written as a single-valued function of
$\bR$ over all of parameter space. Simon \cite{sim} showed that the quantum
adiabatic theorem has a line bundle structure inherent in it, and that
Schrodinger's equation defines a parallel transport of the quantum state
around the line bundle. Berry's phase is the signature that the associated
connection has non-vanishing curvature. In this paper we will consider
Berry's original scenario for a two-level system (2LS), though we will remove 
the adiabatic restriction on the environment. Our goal is to obtain 
the corrections to Berry's phase produced by non-adiabatic effects.

The organization of this paper is as follows. In Section~\ref{sec2}, we
introduce a non-perturbative method for determining all non-adiabatic
corrections to Berry's phase. From the derivation, it
will be clear that the effect of transitions between the two energy levels
has been included to all orders in the non-adiabatic coupling. 
The problem of determining these corrections is reduced to solving an
ordinary differential equation, to which numerical methods should provide
solutions in a variety of situations. In Section~\ref{sec3} we work out a 
particular example in great detail. In Section~\ref{sec3a} we apply our method
to this example and determine exactly the non-adiabatic corrections to 
Berry's phase. As a test of our method, 
in Section~\ref{sec3b} we solve the Schrodinger equation exactly 
using a rotating frame transformation, and
use this solution to independently obtain the non-adiabatic corrections to 
Berry's phase. Comparison with the result obtained in Section~\ref{sec3a} 
shows that both methods yield the same result. In Section~\ref{sec3c} we
examine the corrections to Berry's phase obtained from our analysis in the
limit of weak non-adiabaticity. We do this numerically and analytically,
and compare our results with a previous result due to Berry. In 
Section~\ref{sec3d}, we discuss a magnetic resonance experiment that 
provides a realization of this particular example, and show how the 
non-adiabatic corrections to Berry's phase can be observed in measurements 
of the transverse magnetization. Finally, we make closing remarks in 
Section~\ref{sec4}.

\section{General Analysis}
\label{sec2}
As mentioned in the Introduction, we consider a 2LS coupled to an
environmental degree of freedom $\bRt = R(t)(\sin\theta\cos\phi ,\;
\sin\theta\sin\phi ,\; \cos\theta )$ with non-adiabatic time dependence.
The coupling is described by the Hamiltonian,
\be
H(t) = \bRt\cdot\mbox{\boldmath $\sigma$} \hspace{0.1in} .
\label{ham}
\ee
We denote the instantaneous eigenstates of $H(t)$ by $|E_{\pm}(t)\rangle$
with corresponding eigenvalues $E_{\pm}(t)=\pm R(t)$.

Because $H(t)$ has non-adiabatic time dependence, transitions are possible
between the two energy levels. Consequently, if we prepare the 2LS in the
negative energy level $|E_{-}(0)\rangle$ initially, there is a finite
probability amplitude $T_{-}(t)$ to find the 2LS in the positive energy
level $|E_{+}(t)\rangle$ at time $t$. $\TM (t)$ is the transition amplitude,
and the subscript indicates that the transition began in the negative energy
level. Similarly, the probability amplitude that the system is found again
in the negative energy level at time $t$ defines the persistence
amplitude $\PM (t)$. The subscript again indicates that the system was
initially in the negative energy level. The amplitudes $\PP (t)$ and
$\TP (t)$ have analogous definitions.

The 2LS dynamics is determined by the propogator $U(t,0)=\exp [\,
-(i/\hbar )\int_{0}^{t}\, d\tau\, H(\tau )]$:
\be
U(t,0)|E_{\pm}(0)\rangle = P_{\pm}(t) |E_{\pm}(t)\rangle +
                            T_{\pm}(t) |E_{\mp}(t)\rangle \hspace{0.1in} .
\label{psit}
\ee
We will determine $\PM (t)$ and $\TM (t)$ below, though our principle
interest is in $\PM (t)$. The following derivation is easily adapted to 
determine $\PP (t)$ and $\TP (t)$, though we will not provide that derivation
here. It proves convenient to write $U(t,0)$ as a $2\times 2$ matrix:
\be
U(t,0) = \sum_{E_{i}(t), E_{j}(0)}\, U_{ij} |E_{i}(t)\rangle \langle E_{j}(0)|
          = \left( \begin{array}{cc}
                          \PP (t) \hspace{0.1in}  & \hspace{0.1in} 
                                                       \TM (t) \\
                          \TP (t) \hspace{0.1in} & \hspace{0.1in} \PM (t)
                         \end{array}  \right) \hspace{0.1in} .
\label{prop}
\ee

To begin, we divide up the time interval $(0,\; t)$ into $n$ shorter time
intervals of duration $\epsilon = t/n$ by introducing intermediate times
$t_{k}= k\epsilon$, $(k= 0, \cdots , n)$. Later we will let $n \rightarrow
\infty$. Clearly, $U(k)=U(t_{k},\; t_{k-1})$ propogates the state over the
k-th sub-interval, and
\be
U(t,0) = U(n)\cdots U(1) \hspace{0.1in} .
\label{prodU}
\ee
$U(k)$ has the same structure as eqn.~(\ref{prop}): 
\be
U(k) = \left( \begin{array}{cc}
                \Delta\PP (k) \hspace{0.1in} & \hspace{0.1in}
                                                  \Delta\TM (k) \\
                \Delta\TP (k) \hspace{0.1in} & \hspace{0.1in} \Delta\PM (k)
              \end{array} \right) \hspace{0.1in} .
\label{infU}
\ee
As the notation implies, $\Delta P_{\pm}(k)$ and $\Delta T_{\pm}(k)$
are the persistence and transition amplitudes corresponding to the
k-th sub-interval. Noting that,
\be
U(k) \approx 1 -\frac{i\epsilon}{\hbar} H(k) 
   + {\cal O}(\epsilon^{2})\hspace{0.1in} ,
  \label{approU}
\ee
\be
\label{eig}
 \langle E_{\pm}(k)| \approx \langle E_{\pm}(k-1)| +\epsilon
                               \frac{\partial}{\partial t}\langle
                                E_{\pm}(k-1)| 
                        +{\cal O}(\epsilon^{2}) \hspace{0.1in} ,
\ee
and using eqns.~(\ref{prop}), (\ref{approU}), and (\ref{eig}), one finds
that,
\be
\Delta P_{\pm}(k) = 1+i\epsilon\dot{\gamma}_{\pm}(k) -\frac{i\epsilon}{\hbar}
                      E_{\pm}(k) \hspace{0.3in} ; \hspace{0.3in}
                       \Delta T_{\pm}(k) = -\epsilon \Gamma_{\pm}(k)
                        \hspace{0.1in} .
\label{spt}  
\ee
Here a dot over a symbol indicates time differentiation, and
\be
i\dot{\gamma}_{\pm}(k) = -\langle E_{\pm}(k)|\dot{E}_{\pm}(k)\rangle
   \hspace{0.3in} ; \hspace{0.3in} \Gamma_{\pm}(k) = 
     \langle E_{\mp}(k)|\dot{E}_{\pm}(k)\rangle \hspace{0.1in} ,
\label{dots}
\ee
with
\bd
\Gamma_{+}(k) = -\Gamma_{-}^{\ast}(k) \hspace{0.1in} .
\ed
$\gamma_{\pm}(k)$ are the Berry phases for the $\pm$ energy levels, 
and $\Gamma_{\pm}(k)$ are known as the non-adiabatic couplings for the
$\pm$ energy levels. 

Inserting eqn.~(\ref{infU}) repeatedly into eqn.~(\ref{prodU}), and
carrying out the necessary matrix multiplications, one can show using
induction that,
\bea
\lefteqn{ \PM (t)  = \prod_{k=1}^{n}\Delta\PM (k) } \nonumber \\
        &   &  \hspace{0.75in}
                + \left[ \prod_{k=n_{1}+1}^{n} \Delta\PM (k) \right] 
                   \left[\Delta\TP (n_{1}) \right] \left[
                    \prod_{k=n_{2}+1}^{n_{1}-1}\Delta\PP (k)\right]
                     \left[ \Delta\TM (n_{2})\right]\left[
   \vspace{0.30in}
                      \prod_{k=1}^{n_{2}-1}\Delta\PM (k)\right] \nonumber \\
        &   &  \hspace{1.60in} + \hspace{0.2in} \cdots  \hspace{0.2in} .
\label{close}
\eea
We can make this equation more intelligible by introducing the amplitude 
${\cal P}_{\pm}(i, j-1)$ to persist {\em without\/} any transitions
in a given energy level over the interval $(t_{j-1},\; t_{i})$:
\bd
{\cal P}_{\pm}(i, j-1) = \prod_{k=j}^{i}\Delta P_{\pm}(k) 
  \hspace{0.1in} .
\ed
Eqn.~(\ref{close}) can then be re-written as,
\be
\PM (t)  =  \cPM (n,0) 
         + \cPM (n,n_{1})\,\Delta\TP (n_{1})\,
                                \cPP (n_{1},n_{2})\,\Delta\TM (n_{2})
                                 \cPM (n_{2},0) 
         + \hspace{0.2in} \cdots \hspace{0.2in} .
\label{close2}
\ee
Each term in eqn.~(\ref{close2}) gives the amplitude that the state of the
2LS follows a particular time sequence that begins and ends in the
negative energy level. For any given time sequence, each sub-interval 
will have an amplitude associated with it which indicates whether a 
transition occurred during it $(\Delta T(k))$, or not
$(\Delta P(k))$. Thus the first term gives the amplitude that the system
undergoes zero transitions; the second term gives the amplitude that two
transitions occurred ( in sub-intervals $n_{1}$ and $n_{2}$).
The remaining terms correspond to 4-transitions, 6-transitions, etc.\ .
Only an even number of transitions are possible since the time development
begins and ends in the negative energy level. Thus each transition out of 
this level must eventually be followed by a transition back into it. One
can set-up a diagrammatic calculus to produce all the terms in 
eqn.~(\ref{close2}),
complete with rules for assigning a probability amplitude to each diagram,
though we won't take the time to work that out here.

Similarly, one can show that
\bea
\lefteqn{ \TM (t)  = \cPP(n,n_{1})\:\Delta\TM (n_{1})\:
            \cPM (n_{1},0) }  \nonumber \\
        &   &  \hspace{0.75in} +\cPP (n,n_{1})\:\Delta\TM (n_{1})\:
                \cPM (n_{1}),n_{2})\:\Delta\TP (n_{2})\:\cPP (n_{2},n_{3})\:
                 \Delta\TM (n_{3})\:\cPM (n_{3},0) \nonumber \\
        &   &  \hspace{1.15in} + \hspace{0.15in} \cdots \hspace{0.15in} .
\label{tclose}
\eea
Here only an odd number of transitions can occur since the 2LS must finish
in the positive energy level after having started in the negative energy 
level. Further simplification is possible if we introduce
\bd
E(n_{1},n_{2}) = \Delta\TP (n_{1})\:\cPP(n_{1},n_{2})\,\cPM^{-1}(n_{1},n_{2})
                  \:\Delta\TM (n_{2}) \hspace{0.1in} ,
\ed
and insert $1=\cPM (k) \cPM^{-1}(k)$ appropriately into eqns.~(\ref{close2})
and (\ref{tclose}). One finds that,
\bd
\PM (t) =\cPM (n,0)\:\left[ 1 + E(n_{1},n_{2}) + E(n_{1},n_{2})E(n_{3},n_{4})
           + \hspace{0.15in} \cdots \hspace{0.15in} \right] \hspace{0.1in} ,
\ed
and,
\begin{eqnarray*}
\TM (t) & = &  \cPM (n,0)\:\left\{ \,\cPP (n,n_{1})\:\Delta\TM (n_{1})\:
                 \cPM^{-1}(n,n_{1}) \,\right\} \nonumber \\
        &   & \hspace{0.6in} \times \:\left[ 1 + E(n_{2}, n_{3}) +
              E(n_{2},n_{3})E(n_{4},n_{5}) + \hspace{0.15in} \cdots 
                \hspace{0.15in} \right] \hspace{0.1in} .
\end{eqnarray*}
Using eqn.~(\ref{spt}), and recalling that $n$ is large, one can show that,
\bd
{\cal P}_{\pm}(n_{1},n_{2}) = \exp\left[\sum_{k=n_{2}}^{n_{1}}\,\epsilon
                               \left\{i\dot{\gamma}_{\pm}(k) -(i/\hbar )
                                E_{\pm}(k)\right\}\,\right] \hspace{0.1in} ,
\ed
and,
\bd
E(n_{1},n_{2}) = - \left[\,\epsilon F^{\ast}(n_{1})\,\right]\,\left[\,
                    \epsilon F(n_{2})\,\right] \hspace{0.1in} ,
\ed
with,
\bd
F(m) = \Gamma_{-}(m) \exp\left[\frac{i}{\hbar}\sum_{k=0}^{m}\,\epsilon\left\{
        (E_{+}(k) - E_{-}(k)) - \hbar (\dot{\gamma}_{+}(k)-
         \dot{\gamma}_{-}(k)) \right\} \right] \hspace{0.1in} .
\ed
So far, we have only considered one particular choice of intermediate times.
We must now sum over all $t_{k}$ (maintaining the proper time orderings).
This yields the following expression for $\PM (t)$:
\be
\PM (t) = \exp\left[\, i\gamma_{-}(t) -\frac{i}{\hbar}\int_{0}^{t}\,
           d\tau\, E_{-}(\tau ) \right]\: S(t) \hspace{0.1in} ,
\label{bigdef}
\ee
where,
\bea
S(t) & = & 1 - \int_{0}^{t}dy_{1}\,F^{\ast}(y_{1})\,\int_{0}^{y_{1}}\,
                dx_{1}\, F(x_{1}) \nonumber \\
     &   & \hspace{0.4in} +\int_{0}^{t}dy_{1}\, F^{\ast}(y_{1})\,
             \int_{0}^{y_{1}}\, dx_{1}\, F(x_{1})\,\int_{0}^{x_{1}}\,
              dy_{2}\, F^{\ast}(y_{2})\,\int_{0}^{y_{2}}\, dx_{2}\, F(x_{2})
               \nonumber \\
     &   &  \hspace{0.8in}   - \hspace{0.2in} \cdots \hspace{0.2in} ,
\label{sdef}
\eea
and,
\be
F(t)=\Gamma_{-}(t)\exp\left[ i\int_{0}^{t}\, d\tau\, \delta (\tau ) \right]
  \hspace{0.2in} ; \hspace{0.2in} \delta (\tau ) = \left[
    \frac{E_{+}(\tau ) - E_{-} (\tau )}{\hbar} - (\dot{\gamma}_{+} (\tau )
      - \dot{\gamma}_{-} (\tau ) ) \right] \hspace{0.1in} .
\ee
We see that $S(t)= A(t) \exp[i\rho (t)]$ contains all the consequences of the
non-adiabatic time dependence, and that it includes transitions between the
levels to all orders in the non-adiabatic couplings $\Gamma_{\pm}(t)$. It
is also clear that $\rho (t)$ contains all the non-adiabatic corrections
to Berry's phase $\gamma_{-}(t)$. We close this Section by presenting a
procedure for evaluating $S(t)$ which promises to be useful in a 
variety of situations.

It is a simple matter to write eqn.~(\ref{sdef}) as an integral equation
for $S(t)$:
\be
S(t) = 1 -\int_{0}^{t}\, dy\, F^{\ast}(y)\,\int_{0}^{y}\, dx \, F(x) S(x)
 \hspace{0.1in} .
\label{inteqn}
\ee
Introducing the auxiliary quantity $I(t)$,
\be
I(t) = \int_{0}^{t}\, dx \, F(x) S(x) \hspace{0.25in} \Longleftrightarrow
  \hspace{0.25in} S(t) = \frac{1}{F(t)}\frac{dI}{dt} \hspace{0.1in} ,
\label{Idef}
\ee
and differentiating eqn.~(\ref{inteqn}) with respect to $t$ yields an
ordinary differential equation for $I(t)$:
\be
\frac{d^{2}I}{dt^{2}} + \left(\frac{\dot{F}}{F}\right)\frac{dI}{dt} +
   |F|^{2} I = 0 \hspace{0.1in} .
\label{ode}
\ee
From eqn.~(\ref{Idef}), the appropriate initial conditions are $I(0) = 0$,
and $\dot{I}(0) = F(0)$ (note that $S(0)=1$ according to eqn.~(\ref{inteqn})).
Determining $I(t)$ reduces to solving eqn.~(\ref{ode}), either numerically
or analytically. This is expected to be possible in a variety of 
situations. From $I(t)$ we determine $S(t)$, and hence,
\be
\tan\rho (t) = \frac{Im\; S(t)}{Re\; S(t)} \hspace{0.25in} ; \hspace{0.25in}
  A(t) = \sqrt{\left( Re\; S(t) \right)^{2} + \left( Im\; S(t) \right)^{2}}
   \hspace{0.1in} .
\label{final}
\ee
$I(t)$ also allows us to express $\TM (t)$ more succinctly. 
Using the above results, one can show that,
\be
\TM (t) = - \exp\left[ i\gamma_{+}(t) - \frac{i}{\hbar}\int_{0}^{t}\,
             d\tau\, E_{+}(\tau ) \right]\: I(t) \hspace{0.1in} .
\label{finalT}
\ee
Eqns.~(\ref{bigdef})---(\ref{finalT}) constitute a general approach for
determining completely the consequences of the non-adiabatic motion of the
environment. Specifically, $\rho (t)$ contains all non-adiabatic corrections 
to Berry's phase, while $A(t)$ describes the reduced amplitude for the 2LS
to be found in the negative energy level at time $t$ due to transitions. 
In the following Section, we examine a particular example which is both 
experimentally realizable, and yet simple enough that our equations can be 
evaluated without approximation, and tested against the exact solution of 
the Schrodinger equation.
    
\section{Analysis of a Particular Example}
\label{sec3}
In this Section we will examine in great detail the interaction of a spin 1/2
with a time-varying magnetic field $\bBt$. The magnetic field is assumed
to precess about the z-axis at a fixed angle $\theta$, at a constant
precession rate $\dot{\phi}(t)=\omega$, and with constant magnitude 
$|\bBt |= B$. Such magnetic fields are encountered regularly in experiments 
involving nuclear magnetic resonance (NMR). As such, the results of this 
Section should be amenable to experimental test. The coupling of a spin to a 
magnetic field is described by the Zeeman Hamiltonian which, for a spin 1/2,
has the same form as eqn.~(\ref{ham}) with the substitution 
$\bRt = -\gamma\hbar\bBt /2$.
Throughout this Section we will stick with the notation of eqn.~(\ref{ham}),
though it is a simple matter to substitute for $\bRt$ when necessary.
We will occasionally refer to $\bRt$ as the magnetic field, though this 
is not literally true.

\subsection{Non-Adiabatic Effects: General Analysis}
\label{sec3a}
In this subsection, we will determine all
non-adiabatic corrections  by evaluating $S(t)$ using the general analysis of 
Section~\ref{sec2}. To begin, we must determine the instantaneous
eigenstates $|E_{\pm}(t) \rangle$ of $H(t)$. Straightforward analysis of
eqn.~(\ref{ham}) gives, 
\be
|E_{+}(t)\rangle = \left(\begin{array}{c}
                            \cos\frac{\theta}{2} \\
                            \sin\frac{\theta}{2} \exp [i\omega t]
                         \end{array} \right) \hspace{0.4in} ;
\hspace{0.4in}
|E_{-}(t)\rangle = \left( \begin{array}{c}
                             \sin\frac{\theta}{2} \\
                             -\cos\frac{\theta}{2}\exp [i\omega t]
                          \end{array} \right) \hspace{0.1in} ,
\label{inste}
\ee
where $\bRt = R(\sin\theta\cos\omega t,\;\sin\theta\sin\omega t,\;
\cos\theta)$. Combining eqn.~(\ref{inste}) with eqn.~(\ref{dots}) gives, 
\be
\left\{ \begin{array}{l}
          \Gamma_{-}(t) = -\frac{i\omega}{2}\sin\theta \equiv -iC \\
          \delta (t) = \frac{2R}{\hbar} -\omega\cos\theta 
       \end{array} \right. \hspace{0.2in} \Longrightarrow \hspace{0.2in}
F(t) = -iC\exp [i\delta t] \hspace{0.1in} . 
\label{specifics}
\ee
$F(t)$ is now inserted into eqn.~(\ref{ode}) to give: 
\bd
\ddot{I} -i\delta\dot{I} + C^{2}I = 0 \hspace{0.1in} .
\ed
This equation is easily solved since it has constant coefficients. One finds,
\be
I(t) = -\frac{i\omega\sin\theta}{\Omega_{0}}\exp\left[\frac{i\delta t}{2}
         \right]
         \sin\left(\frac{\Omega_{0}t}{2}\right) \hspace{0.1in} ,
\label{explI}
\ee
with $\Omega_{0} = \sqrt{\delta^{2} + 4C^{2}}$. $S(t)$ follows from
eqn.~(\ref{Idef}): 
\bea
Re\; S(t) & = & \cos\frac{\delta t}{2}\cos\frac{\Omega_{0}t}{2}
                   + \cos\Delta\theta\sin\frac{\delta t}{2}
                      \sin\frac{\Omega_{0} t}{2} 
\label{Re2} \\
Im\; S(t) & = & -\sin\frac{\delta t}{2}\cos\frac{\Omega_{0}t}{2} +
                  \cos\Delta\theta\cos\frac{\delta t}{2}
                   \sin\frac{\Omega_{0}t}{2} \hspace{0.1in} .
\label{Im2}
\eea
Here $\cos\Delta\theta = \delta /\Omega_{0}$: the physical significance
of $\Delta\theta$ will become clear in the following subsection. 

\subsection{Non-Adiabatic Effects: Rotating Frame Analysis}
\label{sec3b}
In this subsection we obtain the exact solution to Schrodinger's equation
for the particular example considered in this Section, and derive from
it $S(t)$. This result will be compared with that obtained in
Section~\ref{sec3a}, thus providing a test of our approach.

The exact soultion can be found using a rotating coordinate frame
analysis. In the lab frame, the Schrodinger equation is,
\be
i\hbar \frac{\partial}{\partial t} |\psi (t) \rangle =
  H(t) |\psi (t) \rangle \hspace{0.1in} ,
\label{schro}
\ee
where $H(t)$ is given by eqn.~(\ref{ham}). We can transform to a frame
that rotates with $\bRt$ using the unitary operator,
\bd
{\cal U}(t) = \exp\left[ -\frac{i\omega t}{2}\sigma_{z}\right] 
                  \hspace{0.1in} .
\ed
Writing $|\psi (t) \rangle = {\cal U}(t) |\overline{\psi}(t)\rangle$, and 
substituting
into eqn.~(\ref{schro}) gives the Schrodinger equation in the rotating
frame:
\bd
i\hbar\frac{\partial}{\partial t} |\overline{\psi} (t)\rangle =
   \overline{H} |\overline{\psi} (t) \rangle \hspace{0.1in} ,
\ed
where,
\bd
\overline{H} = {\cal U}^{\dagger}H{\cal U}-i\hbar {\cal U}^{\dagger}
   \dot{{\cal U}}=
  \left( \begin{array}{cc}
            \left\{ R\cos\theta - \frac{\hbar\omega}{2}\right\}
               & R\sin\theta \\
            R\sin\theta & -\left\{ R\cos\theta -\frac{\hbar\omega}{2}\right\}
         \end{array} \right) \hspace{0.1in} .
\ed
$\overline{H}$ is clearly time-independent, as expected, since the magnetic
field is stationary in the rotating frame. The z-component of the magnetic
field has been altered by the transformation. The magnetic field now makes 
an angle $\overline{\theta}$ with the z-axis given by,
\be
\tan\overline{\theta} = \frac{R\sin\theta}{R\cos\theta -
                          \frac{\hbar\omega}{2}} \hspace{0.1in} .
\label{tbar}
\ee
Note that $\overline{\phi}=0$ since $\overline{H}$ is real. 

Being time-independent, $\overline{H}$ has stationary states. The energies are
\be
\overline{E}_{\pm} = \pm\sqrt{\left( R\cos\theta - \frac{\hbar\omega}{2}
                       \right)^{2} + R^{2} \sin^{2}\theta} =
                    \pm \frac{\hbar\Omega_{0}}{2} \hspace{0.1in} ,
\label{rote}
\ee
and $\Omega_{0}$ was defined in Sec.~\ref{sec3a}. The eigenstates are:
\be
|\overline{E}_{+}\rangle = \left( \begin{array}{c}
                                     \cos\frac{\overline{\theta}}{2} \\
                                     \sin\frac{\overline{\theta}}{2}
                                  \end{array} \right) \hspace{0.4in} ;
  \hspace{0.4in} |\overline{E}_{-}\rangle = 
                    \left( \begin{array}{c}
                              \sin\frac{\overline{\theta}}{2} \\
                              -\cos\frac{\overline{\theta}}{2}
                           \end{array} \right) \hspace{0.1in} . 
\label{rotst}
\ee
The initial condition in the lab frame is $|\psi(0)\rangle =
|E_{-}(0)\rangle$, and $|E_{-}(0)\rangle$ is given in eqn.~(\ref{inste}). 
Since $U(0)=1$, the initial condition in the rotating frame is
$|\overline{\psi}(0)\rangle = |E_{-}(0)\rangle$. Expanding $|\overline{\psi}
(0)\rangle$ in the basis $|\overline{E}_{\pm}\rangle$ gives, 
\be
|\overline{\psi}(0)\rangle = a_{+}|\overline{E}_{+}\rangle + a_{-}
                               |\overline{E}_{-}\rangle \hspace{0.1in} ,
\label{inconrot}
\ee
and application of the initial condition gives,
\be
a_{+} = -\sin\frac{\Delta\theta}{2} \hspace{0.4in} ; \hspace{0.4in}
  a_{-} = \cos\frac{\Delta\theta}{2} \hspace{0.1in} .
\label{apm}
\ee
Here $\Delta\theta\equiv\overline{\theta}-\theta$, and is the same
$\Delta\theta$ as appeared in Sec.~\ref{sec3a}. One can see this by
using eqn.~(\ref{tbar}) and standard 
trigonometric identities to show that $\cos\Delta\theta =
\delta /\Omega_{0}$, just as we found for $\Delta\theta$ in Sec.~\ref{sec3a}. 
Physically, $\Delta\theta$
is the change in the angle the magnetic field makes with the z-axis, as
seen in the rotating and lab frames. 
From 
eqn.~(\ref{inconrot}) we can immediately write,
\bd
|\overline{\psi}(t)\rangle = a_{+}\exp\left[ -\frac{i\Omega_{0}t}{2}\right]
                               |\overline{E}_{+}\rangle + a_{-}
                              \exp\left[ \frac{i\Omega_{0}t}{2}\right]
                                 |\overline{E}_{-}\rangle \hspace{0.1in} .
\ed
Transforming back to the lab frame gives the exact solution 
$|\psi (t)\rangle$ :
\bea
\lefteqn{ {} \hspace{-0.in} 
           |\psi (t) \rangle =  \exp\left[ -\frac{i\omega t}{2}\right] }
 \nonumber \\ 
   & & {} \hspace{0.6in}
  \times \left\{ \begin{array}{l}
                    a_{+}\exp\left[ -\frac{i\Omega_{0}t}{2}\right]
                      \left( \begin{array}{c}
                            \cos\frac{\overline{\theta}}{2} \\
                            \sin\frac{\overline{\theta}}{2}\exp [i\omega t]
                         \end{array} \right) 
                       + a_{-}\exp\left[\frac{i\Omega_{0}t}{2}\right]
                 \left( \begin{array}{c}
                           \sin\frac{\overline{\theta}}{2} \\
                           -\cos\frac{\overline{\theta}}{2}\exp [i\omega t]
                        \end{array} \right) 
                 \end{array} \right\} \hspace{0.1in} .
\label{exacts}
\eea
From eqn.~(\ref{exacts}) we can obtain the persistence amplitude
$P_{-}(t)=\langle E_{-}(t)|\psi (t)\rangle $. Using eqn.~(\ref{inste}), 
we find,
\bd
P_{-}(t) = \exp\left[ -\frac{i\omega t}{2} \right]
            \left\{ a_{+}^{2}\exp\left[ -\frac{i\Omega_{0}t}{2} \right]
                     + a_{-}^{2}\exp\left[\frac{i\Omega_{0}t}{2}\right]
             \right\} \hspace{0.1in} .
\ed
This expression can be straightforwardly re-written as,
\bd
P_{-}(t) = \exp\left[ i\gamma_{-} - \frac{i}{\hbar}\int_{0}^{t}\,
             d\tau\, E_{-}(\tau ) \right]
             \left\{ \exp\left[ -\frac{i\delta t}{2}\right]
               \left( a_{+}^{2}\exp\left[ -\frac{i\Omega_{0}t}{2}\right]
                       + a_{-}^{2}\exp\left[\frac{i\Omega_{0}t}{2}\right]
                 \right) \right\} \hspace{0.1in} .
\ed
Thus, the factor in curly brackets is $S(t)$ (see eqn.~(\ref{bigdef})),
as determined from the exact solution of Schrodinger's equation. The exact 
solution thus yields,
\bea
Re\; S(t) & = & \cos\frac{\delta t}{2}\cos\frac{\Omega_{0}t}{2}
                 +\cos\Delta\theta\sin\frac{\delta t}{2}
                   \sin\frac{\Omega_{0}t}{2} \label{reS2} \\
Im\; S(t) & = & -\sin\frac{\delta t}{2}\cos\frac{\Omega_{0}t}{2} +
                  \cos\Delta\theta\cos\frac{\delta t}{2}
                    \sin\frac{\Omega_{0}t}{2} \label{imS2} \hspace{0.1in} .
\eea
Comparing eqns.~(\ref{reS2}) and (\ref{imS2}) with eqns.~(\ref{Re2}) and
(\ref{Im2}), we see that our approach gives precisely the same result for
$S(t)$ as the exact solution of the Schrodinger equation.

\subsection{Non-Adiabatic Effects: Numerical and Analytical Evaluation}
\label{sec3c}
Here we explicitly evaluate the non-adiabatic corrections to Berry's phase.
The exact result (including all non-adiabatic corrections) will be
evaluated numerically. We also determine an analytical approximation for 
these corrections 
in the limit of weak non-adiabaticity. We shall see that the analytic
approximation agrees quite well with the exact result in this limit. We
also compare our results with an earlier result due to Berry.

We begin by substituting eqns.~(\ref{Re2}) and (\ref{Im2}) into 
eqn.~(\ref{final}). This yields, 
\be
\tan\rho = \frac{\terma -\termb}{1+\terma\termb} \hspace{0.1in} ,
\label{tanrho}
\ee
where $g\equiv\cos\Delta\theta = \delta /\Omega_{0}$. Introducing
$\varepsilon$ through the relation,
\be
\tan\frac{\varepsilon t}{2} = g\tan\frac{\Omega_{0}t}{2} \hspace{0.1in} ,
\label{def2}
\ee
eqn.~(\ref{tanrho}) becomes,
\bd
\tan\rho = \tan\left[ \frac{(\varepsilon - \delta )t}{2} \right] 
   \hspace{0.1in} ,
\ed
so that
\be
\rho = \frac{(\varepsilon -\delta )}{2}t + n\pi \hspace{0.1in} .
\label{rhor}
\ee
We focus on the $n=0$ branch. Clearly, we must determine $\varepsilon$.
To do this, we differentiate eqn.~(\ref{def2}) with respect to $\omega$
at fixed $t$. This gives:
\be
\frac{d\varepsilon}{d\omega} =
  \frac{2\cos^{2}\frac{\varepsilon t}{2}}{t}
   \left[ \frac{dg}{d\omega}\tan\frac{\Omega_{0}t}{2} +
    \frac{gt}{2\cos^{2}\frac{\Omega_{0}t}{2}} \frac{d\Omega_{0}}{d\omega}
     \right] \hspace{0.1in} .
\label{ode1}
\ee
The appropriate initial condition for eqn.~(\ref{ode1}) is,
\bd
\varepsilon (\omega = 0) = \Omega_{0} (\omega = 0) = \frac{2R}{\hbar}
  \hspace{0.1in} ,
\ed
which follows from eqn.~(\ref{def2}), since $g(\omega = 0) =1$. It proves
useful to write eqn.~(\ref{ode1}) in dimensionless form. For this purpose,
we introduce the following definitions:
\be
x=\frac{\hbar\omega}{2R} \hspace{0.4in} ; \hspace{0.4in}
 \tau = \left(\frac{2R}{\hbar}\right) t \hspace{0.4in} ; \hspace{0.4in}
  \epsilon = \left( \frac{\hbar}{2R}\right)\varepsilon \hspace{0.1in} .
\label{def3}
\ee
It also proves useful to define dimensionless versions of $\delta$ and
$\Omega_{0}$ (see Sec.~\ref{sec3a}):
\bea
d & =  \left(\frac{\hbar}{2R}\right)\delta & = 1-x\cos\theta 
   \label{ddef} \\
e & = \left(\frac{\hbar}{2R}\right)\Omega_{0} & = \sqrt{1-2x\cos\theta
        +x^{2}} \label{edef} \hspace{0.1in} .
\eea
Thus, $g=\delta /\Omega_{0}= d/e$. We will integrate eqn.~(\ref{ode1}) 
over the range $0\le \omega\le\omega_{f}$, and will use $T_{f} =
2\pi /\omega_{f}$ to define the time scale ($t=(2\pi /\omega_{f})s$).
Then,
\bd
\tau = \frac{2\pi s}{x_{f}} \hspace{0.1in} .
\ed
With all these definitions in place, eqn.~(\ref{ode1}) takes the following
dimensionless form:
\be
\frac{d\epsilon}{dx} = \frac{x_{f}\cos^{2}\frac{\pi s\epsilon}{x_{f}}}{s\pi}
                        \left[ \frac{dg}{dx}\tan\frac{\pi se}{x_{f}}
                         +\frac{\pi sg}{x_{f}\cos^{2}\frac{\pi se}{x_{f}}}
                           \frac{de}{dx} \right] \hspace{0.1in} .
\label{ode2}
\ee
Numerical integration of this equation yields $\epsilon (x,\; t)$, which
in turn gives $\rho (t)$ via eqns.~(\ref{def3}) and (\ref{rhor}). 
We remind the reader that this numerical result contains all non-adiabatic
corrections to Berry's phase. The numerical result for $\rho$ is given in
Figure~1 (curve A). The integration was done with $s=1$ (one
precession cycle), $\theta = 60^{\circ}$, and $x_{f}=0.3$.

It is possible to find a simple analytical expression for $\rho (t)$ when
$x \ll 1$. In this case, eqn.~(\ref{ode2}) can be treated iteratively.
We only carry out the first step of this iteration procedure since a
numerical analysis is better suited to handle the case when $x$ is not
small. Noticing that $\epsilon (0) = e(0)$, and since we are assuming that
$x\ll 1$, the first iteration step substitutes $\epsilon_{0}(x) = e(x)$ into
the right-hand side (RHS) of eqn.~(\ref{ode2}). Solving the resulting
differential equation yields $\epsilon_{1}$, the improved $\epsilon$,
which serves as the input for the second iteration step. Thus, $\epsilon_{1}$
is plugged into the RHS of eqn.~(\ref{ode2}) which is solved to yield
$\epsilon_{2}$, etc.. Carrying out the first iteration yields the following
differential equation for $\epsilon_{1}$:
\bd
\frac{d\epsilon_{1}}{dx} = g\frac{de}{dx} + \frac{x_{f}
    \sin\frac{2\pi s}{x_{f}}}{2\pi s}\frac{dg}{dx} \hspace{0.1in} .
\ed
Since $x_{f}\ll 1$, the sine function on the RHS oscillates rapidly, and
so the second term on the RHS is not expected to contribute significantly
to $\epsilon_{1}$. We will see below that this is, in fact, the case. Thus 
we drop the second term and integrate the resulting equation to get,
\bd
\epsilon_{1}(x) = 1 + \int_{0}^{x_{f}}\, dx\, g\frac{de}{dx} \hspace{0.1in}
  .
\ed
Since $x\ll 1$, we can evaluate the integrand to second order in $x$ using
eqns.~(\ref{ddef}) and (\ref{edef}), and then carry out the integration. 
This gives, 
\bd
\epsilon_{1}(x) = 1-x\cos\theta + \frac{x^{2}}{2}\sin^{2}\theta +
  \frac{2x^{3}}{3}\sin^{2}\theta\cos\theta + \cdots \hspace{0.2in} .
\ed
From this we find,
\be
\rho_{1}(t) = \omega t \left[ c_{1}x\sin^{2}\theta + c_{2}x^{2}\sin^{2}\theta
                 \cos\theta + \cdots \hspace{0.2in} \right] \hspace{0.1in} ,
\label{anlr}
\ee
where $c_{1} = 1/4$, and $c_{2} = 1/3$. Eqn.~(\ref{anlr}) is plotted in
Figure~1 (curve B). We see that it agrees quite well with the
numerical evaluation of the exact result (to within $1\%$) for the range
of x-values considered. As anticipated, we see from Figure~1 that
the discarded oscillatory term does not contribute significantly to
$\rho$ when $x\ll 1$.

Berry has worked out an adiabatic iteration procedure which generates an
asymptotic expansion for corrections to Berry's 
phase \cite{br2}. The procedure is adiabatic in that it ignores non-adiabatic
transitions in all orders of iteration. This procedure yields a 
sequence of phase approximants, with each approximant containing powers of $x$
to infinite order. Each iteration renormalizes the coefficients
of the powers of $x$ obtained in the previous iteration step.
Berry has shown that the optimum number of terms to keep in the asymptotic
expansion is $n\sim 1/x$. The sequence of phase approximants initially
improves with successive iterations, though ultimately, the sequence
diverges because of transitions introduced by the non-adiabatic time
dependence.

Berry applied this procedure to the example we are considering
in this Section. He carried out the first iteration step 
and worked out the resulting corrections to Berry's phase to order $x^{2}$.
His result has the same
functional form as eqn.~(\ref{anlr}), though he finds $c_{1}=1/2$,
and $c_{2}=1$. We believe that the discrepancy with our values 
for $c_{1}$ and $c_{2}$ arises from truncation of the adiabatic iteration 
procedure at the first step. Such a truncation of the asymptotic expansion
produces a non-optimum approximation for $\rho (t)$ when $x\ll 1$
(see remark above), and
consequently, non-optimum values for $c_{1}$ and $c_{2}$.
It is clear that Berry's intention was to illustrate his method; we
believe that if a sufficient number of iteration steps were carried out, the
two approaches would produce equivalent values for $c_{1}$ and $c_{2}$.  
The single-iteration result is plotted as curve C in Figure~1.

\subsection{Experimental Realization: Nuclear Magnetic Resonance}
\label{sec3d}
One of the first observations of Berry's phase was by Suter et.\ al.\ 
\cite{sut} using nuclear magnetic resonance (NMR). In this experiment,
the rotating magnetic field precessed about the z-axis in the 
manner assumed in this Section. Measurement of the transverse
magnetization $\langle M_{\perp}(t)\rangle$ allowed observation of Berry's
phase. We now show that this same measurement (not so surprisingly)
will also reveal the non-adiabatic corrections to Berry's phase determined 
above. 

If initially the spin 1/2 has a component transverse to
$\bB (0)$, the spin will begin to precess about $\bB (0)$. If $\bBt$
does not evolve too rapidly, the spin precession simply follows $\bBt$.
To simplify the analysis, we assume,
\bd
|\psi (0)\rangle = \frac{1}{\sqrt{2}}\left[ |E_{+}(0)\rangle +
    |E_{-}(0)\rangle \right] \hspace{0.1in} ,
\ed
corresponding to the spin being aligned initially along the x-axis in
the lab frame. Using $|\psi (t)\rangle = U(t,\; 0)|\psi(0)\rangle$, and
eqn.~(\ref{psit}), we have,
\bd
|\psi (t)\rangle = \frac{1}{\sqrt{2}}\left[ \;\left\{ \PP (t) + \TM (t)
                      \right\}
    |E_{+}(t)\rangle + \left\{ \PM (t)+\TP (t)\right\} |E_{-}(t)\rangle
       \;\right] \hspace{0.1in} .
\ed

The transverse magnetization $\langle M_{\perp}(t)\rangle = \langle
M_{x}(t) +iM_{y}(t)\rangle$ is given by
\be
\langle M_{\perp}(t)\rangle = {\rm Tr}\:\rho_{d}(t)\left\{\gamma\hbar I^{+}
  \right\} \hspace{0.1in} .
\label{tranmag}
\ee
Here $\rho_{d}(t)=|\psi (t)\rangle \langle \psi (t)|$ is the density
matrix; $I^{+} = I_{x}+iI_{y}$ is the raising operator for angular momentum;
and $\gamma$ is the gyromagnetic ratio. We assume that $t=2\pi /\omega$ so
that $|E_{\pm}(2\pi /\omega )\rangle =|E_{\pm}(0)\rangle$. Also, in the
basis $|E_{\pm}(0)\rangle$, $I^{+} = |E_{+}(0)\rangle\langle E_{-}(0)|$.
Using these results in eqn.~(\ref{tranmag}) one finds,
\bea
\langle M_{\perp}(2\pi /\omega )\rangle & = & \frac{\gamma\hbar}{2}
                                               \left(\PM +\TP\right)
                                                \left(\PP^{\ast}+\TM^{\ast}
                                                 \right) \nonumber \\
 & = & \frac{\gamma\hbar}{2}\left[ \PM\PP^{\ast} + \TP\PP^{\ast} +
           \PM\TM^{\ast} + \TP\TM^{\ast} \right] \hspace{0.1in} .
\label{tranmag2}
\eea
We have already evaluated $\PM$. $\TM$ can be evaluated using 
eqns.~(\ref{explI}) and (\ref{finalT}). $\PP$ and $\TP$ are determined
by suitably adapting the analyses for $\PM$ and $\TM$. 
One finds:
\be
\PM = A\exp\left[ \frac{iRt}{\hbar}-\frac{i\omega t}{2}(1+\cos\theta )
      +i\rho \right] \hspace{0.1in}  ;  \hspace{0.1in} 
                            \PP = A\exp\left[ -\frac{iRt}{\hbar}-
                                  \frac{i\omega t}{2}(1-\cos\theta ) -i\rho
                                   \right] 
\ee
and
\be
\TM = -iC\exp\left[ -\frac{i\omega t}{2}\right] \hspace{0.7in} 
     ;  \hspace{0.7in} \TP = \TM
  \hspace{0.1in} .
\label{nogeo}
\ee
$A$ is determined from eqn.~(\ref{final}), and $C$ from eqn.~(\ref{specifics}). 

Using these results in eqn.~(\ref{tranmag2}) gives,
\be
\langle M_{\perp}(2\pi /\omega )\rangle = \frac{\gamma\hbar}{2} A^{2}
   \exp\left[ i\delta t + i\omega t\left(\frac{x}{2}\sin^{2}\theta +
    \frac{2x^{2}}{3}\sin^{2}\theta\cos\theta\right)\right] +
     \frac{\gamma\hbar}{2}C^{2} \hspace{0.1in} .
\label{tranmag3}
\ee
We have assumed $x\ll 1$ so that we can use $\rho_{1}(t)$ from 
Sec~\ref{sec3c} to give an analytic approximation for $\rho (t)$. In this
limit, one can show that $C^{2}$ is of order $x^{2}$, and so the second term 
on the RHS of eqn.~(\ref{tranmag3}) is negligible compared to the first.
Thus,
\be
\langle M_{x}(2\pi n/\omega ) \rangle = \frac{\gamma\hbar A^{2}}{2}
  \cos\left[ 2\pi n\left(\frac{1}{x} -\cos\theta +
   x^{2}\sin^{2}\theta \left(\frac{1}{2}+\frac{2}{3}x\cos\theta\right)\right)
    \right] \hspace{0.1in} .
\label{nadmag}
\ee
Here $x=\omega /\gamma B$ when we substitute for $R$.
The first two terms in the argument of the cosine function were observed
by Suter et.\ al.\ \cite{sut}. It would be interesting if this experiment
could be repeated to look for the non-adiabatic corrections given in
eqn.(\ref{nadmag}).
 
\section{Closing Remarks}
\label{sec4}
In this paper we have presented a non-perturbative method for determining
all non-adiabatic corrections to Berry's phase. The problem of
determining these corrections has been reduced to solving an ordinary 
differential equation (ODE) for which numerical methods should provide 
solutions in a variety of situations.

We applied our method to a particular example which can be realized as an 
NMR experiment, and whose Schrodinger equation can be solved exactly.
For this example, our method could also be implemented exactly, and we 
saw that it yielded
non-adiabatic corrections which were identical to those obtained from the
exact solution. The non-adiabatic corrections to Berry's phase were evaluated 
numerically, and an analytical approximation was also obtained in the limit of 
weak non-adiabaticity. Both results were compared
with a previous result by Berry. We also discussed how the non-adiabatic
corrections to Berry's phase could be measured using NMR.

We close with some final comments. (1) We stress that our method is
non-perturbative. The object determined by the previously mentioned
ODE contains non-adiabatic
corrections to all orders in the non-adiabatic coupling. (2) The phase
we determine is different from the Aharonov-Anandan phase 
\cite{a+a}. In the scenario we consider, it is the system Hamiltonian
which executes a cyclic evolution. Because the time dependence is 
non-adiabatic, the quantum system does not return to its initial state
at the end of a cycle of the Hamiltonian, and so its state will not, 
in general, 
execute a cyclic evolution. The phase we have evaluated is, in fact,
the Pancharatnam phase \cite{pan,br3,s+b}, and we have explicitly seen
that it reduces to Berry's phase in the limit where the non-adiabaticity
goes to zero. We have also seen, for the example considered in 
Section~\ref{sec3}, that no geometric phase appears in the transition
amplitude (see eqn.~(\ref{nogeo})), in agreement with Berry \cite{br4} 
since $\phi(t)$ is an odd function of $t$ in this case.
(3) It would be interesting to apply the method presented here to the
case of an environment undergoing non-adiabatic stochastic motion.
As discussed in Ref.~\cite{me}, the results of such an analysis should 
impact an ongoing controversy connected with the motion of vortices in
superconductors. The controversy centers around whether certain Berry phase 
effects will be masked
by a secondary process (connected with quasiparticle states bound to the
vortex core) whose activation requires sufficiently large temperature
and/or impurity concentration (see Ref.~\cite{me} for further discussion
and references). We hope to report on this application in a future paper. 

\section*{Acknowledgments}
It is a pleasure to thank Alan Bishop and the T-11 group at Los Alamos
National Laboratory for the hospitality and support they provided during
the time in which this work was done. I would also like to thank 
T. Howell~III for continued support. 

\pagebreak

\section*{Figure Caption}
Figure 1: A plot of the non-adiabatic corrections to Berry's phase $\rho$
versus $x=\hbar\omega /2R$. Curve A is the numerical integration of 
eqn.~(\ref{ode2}) which includes the effects of transitions between
the two energy levels to all orders in the non-adiabatic coupling. 
In this calculation, $s=1$, $\theta =60^{\circ}$, and
$x_{f}=0.3$. Curve B is our analytical approximation for $\rho$. 
Curve C is the first-iteration result of Ref.~\cite{br2}.

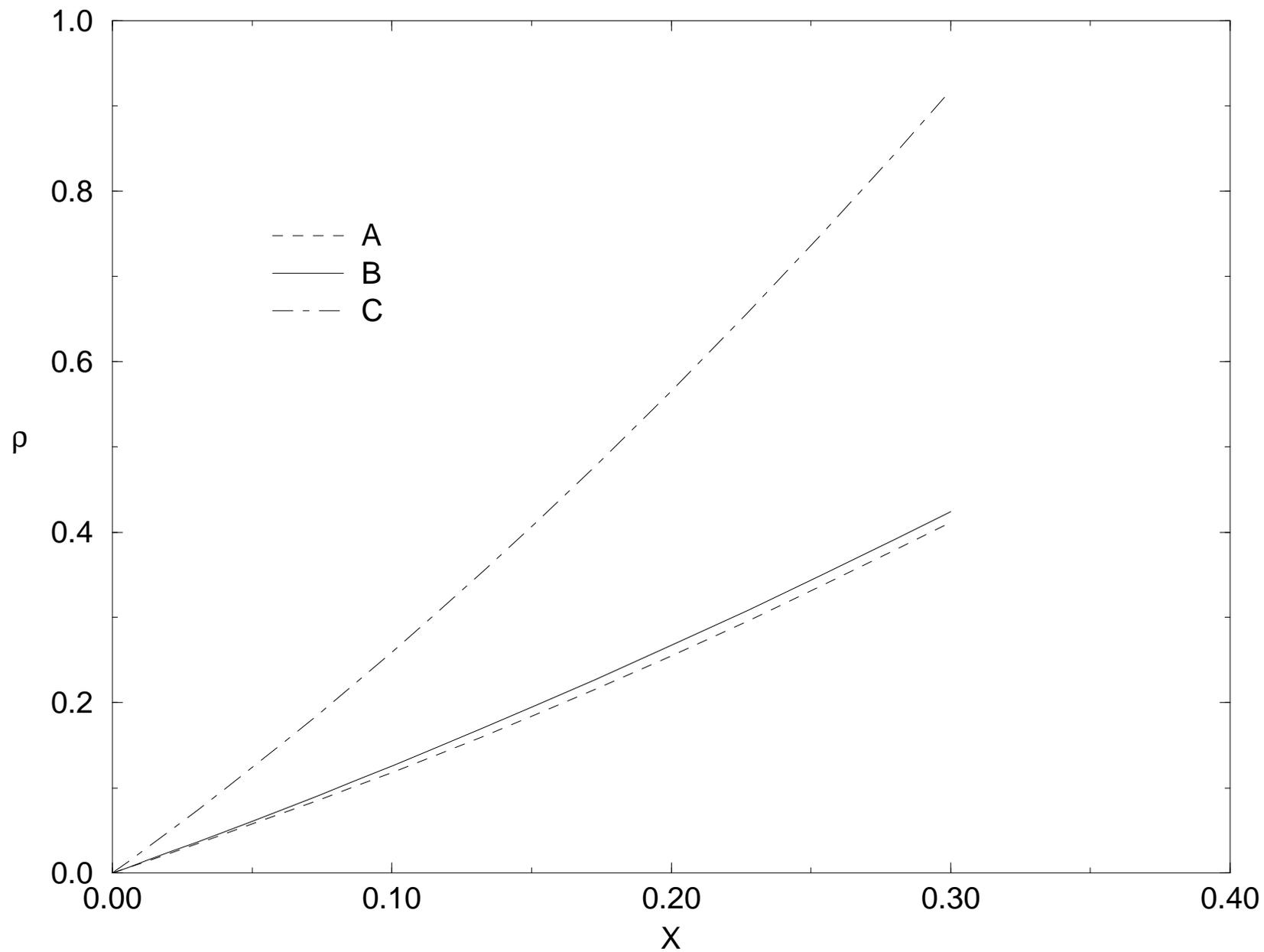
\end{document}